\documentclass{article}

\usepackage{bbm}

\usepackage{graphicx}
\usepackage{amsmath}
\usepackage{amssymb}
\usepackage{amsthm}

\usepackage[update,prepend]{epstopdf}

\usepackage{booktabs}

\newcommand{\ra}[1]{\renewcommand{\arraystretch}{#1}}

\newtheorem*{theorem*}{Theorem}

\title{Asynchronous stochastic price pump}

\author{Misha Perepelitsa and Ilya Timofeyev\footnote{misha@math.uh.edu; ilya@math.uh.edu, 
University of Houston
PGH 631
4800 Calhoun Rd. 
Houston, TX
USA}}

\begin{document}
\maketitle

\begin{abstract}
We propose a model for equity  trading in a population of agents where each agent acts to achieve his or her target stock-to-bond ratio, and, as a feedback mechanism, follows a market adaptive strategy. In this model only a fraction of agents participates in buying and selling stock during a trading period, while the rest of the group accepts the newly set price.

Using numerical simulations we show that the stochastic process settles on a stationary regime for the returns. The mean return can be greater or less than the return on the bond and it is determined by the
parameters of the adaptive mechanism.
When the number of interacting agents is fixed, the distribution of the returns follows the log-normal density.
In this case, we give an analytic formula for the mean rate of return in terms of the rate of change of agents' risk levels and confirm the formula by numerical simulations. 
However, when the number of interacting agents per period is random, the distribution of returns can significantly 
deviate from the log-normal, especially as the variance of the distribution for the number of interacting agents increases.

\end{abstract}

\begin{section}{Introduction}

Human behavior plays a major role in the dynamics of markets. Many market singularities can be attributed to the actions of  investors who make decisions  based on simple investment schemes and do not take into account the cumulative effect they may produce. 

Psychological and sociological aspects such as bias, emotions, and social pressure  admittedly contribute significantly into the investors behavior, quite often  in contrast to what a rational investor might do in a similar situation.

One of the approaches to understanding the effects of complex human behavior on markets is the microscopic simulation of artificial markets. By prescribing a specific type of behavior to different groups of traders and the rules of interaction between traders one can trace their effect on  price dynamics. This approach has been implemented by many authors in a variety of situations. Here we mention just few contributions. Kim-Markowitz \cite{KM} constructed  an agent-based  model for the stock price dynamics in a population of portfolio re-balancers and portfolio insurers that was designed to show the relation between market volatility and the size of the group of the portfolio insurers.   A microscopic model for trading of risky asset, developed in Levy-Levy-Solomon \cite{LLS1, LLS}, considers the interactions of groups of different sizes  of ``chartists'' and ``fundamentalists.'' It was shown that the presence of chartists results in persistent deviations of the stock price from its fundamental value, in a series of speculative bubbles. The model was further investigated  by Cordier-Pareschi-Toscani \cite{CPT}. In fact, many statistical properties of the dynamics of stock prices can be explained as a result of interaction of groups of traders who adopt  different investment strategies, as was shown in  Egenter-Lux-Stauffer \cite{ELS}, Lux-Marchesi \cite{LM}. Models  that take into account herding behavior of traders and the positive feedback they produce on the prices were constructed in Sornette-Andersen \cite{SA}.

Typically, the emphasis of the research is on understanding the critical events in the dynamics of stock prices such as booms, bursts, crashes, super-exponential growth and excessive volatility.

In this paper we focus  on  the trader's behavioral traits that can generate {\it stable}, or more precisely, {\it stationary}  price dynamics patterns -- the kind of dynamics one can see in  a ``typical financial  chart'' in the periods between major changes in the market. We consider adaptive market behavior of traders, as it appears to be the ubiquitous behavioral paradigm that can be found in  many biological systems.

To model an adaptive feedback mechanism  we assume that each agent acts (buys or sells) on the basis of his or her personal experience in the market, as measured by the stock-to-bond ratio of his or her portfolio. The stock-to-bond ratio is one of the most commonly used characteristics of investment portfolios. Here we envision traders who might use technical, sophisticated market analysis to allocate funds between stocks, but who, from time to time, evaluate the total balance of risky versus safe assets and re-adjust the stock-to-bond ratio of the portfolio. In this approach the agents' estimates of the market are strictly subjective. The same event, for example, an emergence and subsequent burst of a price bubble, is evaluated differently by an agent who was lucky to profit from the bubble and the agent who lost money.

In this paper we explore the question if the  investors, acting  strictly in self-interest and using  the stock-to-bond ratio as the only  quantitative characteristic, can create price dynamics with emergent global patterns, such as positive returns on a risky  asset.  To this end we build an agent-based model for trading in which the agents determine the price and adjust their preferences accordingly.

The rate of return on a risk-free investment (bond) is the only exogenous parameter in the model. The feedback mechanism reflects agents' forecast for the upcoming changes in the stock market and specifies how the agent will update his or her stock-to-bond ratio if the market over- or under-performs. Each agent  has the opportunity to change his or her stock-to-bond ratio on average a few times a year, whereas the stock price is updated over short time intervals  (few hundred times  a year). The stock price is determined by the stock supply-demand balance for a randomly selected group of agents. The non-trivial price dynamics emerging from such interactions is the reaction of agents to the changes of the price resulted from other agents behavior, when the whole group shares  a common belief that it is beneficial to have a proportion of wealth invested in stocks.

The paper is organized as follows. In section \ref{sec:2} we introduce a deterministic two-agent model, which illustrates the mechanism of interaction and price formation. Stochastic N-agent model is described in section \ref{sec:3}. This section contains the main finding of the paper: the relation between the mean rate of return of the stock investment and the mean change of agent riskiness.  Section \ref{sec:4} describes the statistical properties of the stochastic process generated by the model. They include the distribution of returns, autocorrelation function and the temporal changes in the first four moments of the distribution of  returns. In the last section we discuss an interpretation of the model as a type of  price bubble.



\end{section}

\begin{section}{Price pump in two-agent model}

\label{sec:2}

We start with the analysis of a deterministic two-agent model which serves as an illustration of the mechanics of its stochastic multi-agent counterpart.  In this model, there are two agents, each described by a state vector of positive numbers
\[
(k_i,\,s_i, b_i) \quad i=1,2, 
\]
where $k_i$-- the agent $i$ preferred stock-to-bond ratio, $\$s_i$ --dollar value of investment in stock and $\$b_i$--investment in bond.
There is a single equity in the market with the current price per stock denoted by $P_0.$ Investment in bonds grows with the gross return rate $r>0$ each trading period.

If ratio $k_i\not=s_i/b_i,$ then agent $i$ is willing to sell or buy stock to bring the ratio to level $k_i.$ We will assume that the stock is infinitely divisible so that any fractional amount of it can be traded. We will assume furthermore that the level $k_i$ is the single parameter that determines the demand of agent $i$ for equity. Under this assumption the demand function is linear in the price of stock denoted by $P.$ That is, at any potential price $P$ at the next period, agent $i$ demands $\$x_i$ value of stock so that
\[
\frac{\frac{P}{P_0}s_i+x_i}{rb_i-x_i}{}={}k_i,
\]
where $x_i$ can be positive (buying) or negative (selling). In this way the demand function $x_i=x_i(P)$ equals
\begin{equation}
\label{eq:s-d}
x_i{}={}-\frac{s_i}{1+k_i}\frac{P}{P_0}{}+{}\frac{rk_ib_i}{1+k_i}.
\end{equation}
Unless $k_1b_i(1+k_2)=k_2b_2(1+k_1)$ and $k_1b_1s_2=k_2b_2s_1,$  there is a single intersection of the  demand curves which determines price $P:$
\[
x_1+x_2{}={}0,
\]
or 
\begin{equation}
\label{Price}
P/P_0{}={}r\left(\frac{k_1b_1}{1+k_1}{}+{}\frac{k_2b_2}{1+k_2}\right)\left(\frac{s_1}{1+k_1}{}+{}\frac{s_2}{1+k_2}\right)^{-1}.
\end{equation}

If both agents have ratios $k_i=s_i/b_i$ at the beginning of trading, then the new price $P=rP_0,$ there is zero amount of stock traded,
and agents retain their investment ratios:
\[
\frac{Ps_i/P_0}{rb_i}{}={}k_i,\quad i=1,2.
\]
We will refer to this kind of process as the steady-state price dynamics. In this scenario the agents have fixed investment ratios, they believe that investing in stock is advantageous (they expect return to be at least $r$), and the stock price, reflecting the agents' belief, grows at the same rate as investment in bonds. The combined wealth of each agent grows as if it were all invested in bond.

When the starting portfolios are out of balance, $k_i\not=s_i/b_i$ then $x_1=-x_2\not=0$ and one agent is buying stock while the other is selling. The outcome is asymmetric, and we assume that it leads to an update of investment ratios from $k_i$ to $\hat{k}_i$ according to the rule:
\begin{equation}
\label{update}
\hat{k}_i{}={}\left\{
\begin{array}{rr}
\alpha k_i, & x_i<0,\\
\beta k_i, & x_i>0,\\
k_i, & x_i=0,
\end{array}
\right.
\end{equation}
where $\alpha,\beta$--positive, same (for simplicity) for both agents.  We designate the range $\alpha>1$ and $\beta<1$ to denote {\it the adaptive feedback mechanism}: the agent increases or decreases his/her riskiness depending if the market is over- or under-performing, as measured by ratios $s_i/b_i.$

 Finally the state variables are set to 
\[
\hat{s}_i{}={}\frac{P}{P_0}s_i+x_i,\quad \hat{b}_i=rb_i-x_i,\quad i=1,2.
\]
Then, the process is iterated. The returns quickly approach the stationary regime, as is illustrated in Figure \ref{fig:2-Return}. The mean return, in general, is different from the safe rate, $r.$  To find the relation between two rates we consider the change of the system over two trading periods. 

Suppose that after one trading period,
\[
\frac{s_i}{b_i}{}={}k_i,\quad i=1,2,
\]
and the ratios are updated so that $\hat{k}_1=\alpha k_1,$ $\hat{k}_2=\beta k_2.$ 
The new price $P$ is set 
\begin{equation}
\label{Price1}
P/P_0{}={}r\left(\frac{\alpha s_1}{1+\alpha k_1}{}+{}\frac{\beta s_2}{1+\beta k_2}\right)\left(\frac{s_1}{1+\alpha k_1}{}+{}\frac{s_2}{1+\beta k_2}\right)^{-1},
\end{equation}
as the intersection of supply-demand curves, see Figure \ref{fig:s-d}.

\begin{figure}[t]
\centering
\includegraphics[width=7cm]{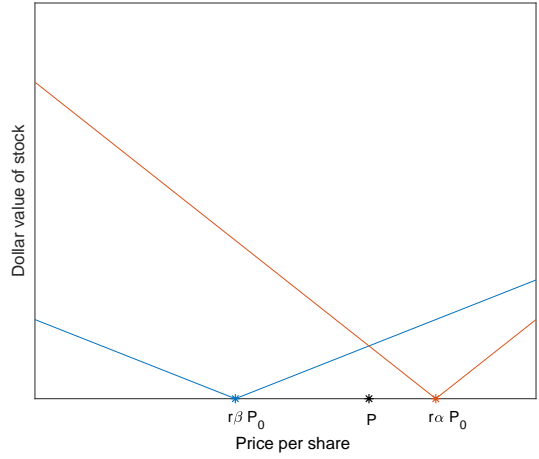}
   \caption{Supply-demand balance. The lines of positive (negative) slope are demand  (supply) curves. The intersection of the supply curve of one agent with the demand of the other determines new price $P.$ }
   \label{fig:s-d}
\end{figure}

This results in reallocation of assets $(\hat{s}_i,\hat{b}_i):$
\[
\frac{\hat{s}_i}{\hat{b}_i}=\hat{k}_i,
\]
with agent 1 buying stocks and agent 2 selling the same amount: the supply-demand curve for agent 1 is to the left of the curve for agent 2. This follows from \eqref{eq:s-d} and formulas for $\hat{k}_i.$  We set $\hat{P}=P$ from \eqref{Price1}. The feedback mechanism gives new investment ratios
\[
\tilde{k}_1=\beta \hat{k}_1,\, \tilde{k}_2=\alpha\hat{k}_2.
\]

In the next trading period new stock price $\tilde{P}$ is determined from the equation similar to $\eqref{Price1}$ with roles of agent 1 and agent 2 reversed. Agent 1 is selling, agent 2 is buying, and  the price changes according to the formula
\begin{equation}
\label{eq:Price2}
\tilde{P}/\hat{P}{}={}r\left(\frac{\beta \hat{s}_1}{1+\beta \hat{k}_1}{}+{}\frac{\alpha \hat{s}_2}{1+\alpha \hat{k}_2}\right)\left(\frac{\hat{s}_1}{1+\beta \hat{k}_1}{}+{}\frac{\hat{s}_2}{1+\alpha \hat{k}_2}\right)^{-1}.
\end{equation}
Recall that $\hat{k}_1=\alpha k_1,$ $\hat{k}_2=\beta k_2,$
\begin{equation}
\label{eq:s1_s2}
\hat{s}_1{}={}\frac{\hat{P}}{P_0}s_1+x_1,\,\hat{s}_2{}={}\frac{\hat{P}}{P_0}s_2-x_1,
\end{equation}
and 
\begin{equation}
\label{eq:x1}
x_1{}={} -\frac{s_1}{1+\alpha k_1}\frac{\hat{P}}{P_0}{}+{}\frac{r\alpha s_1}{1+\alpha k_1}
\end{equation}
From formulas \eqref{Price1}, \eqref{eq:Price2}, \eqref{eq:s1_s2}, \eqref{eq:x1} we obtain  the return over two periods, 
$
\frac{\tilde{P}}{P_0}{}={}\frac{\tilde{P}}{\hat{P}}\frac{\hat{P}}{P_0}
$
as a function of $(k_1,k_2,s_1,s_2).$   With this, we define function
\begin{equation}
\label{amplification}
A(k_1,k_2,s_1,s_2){}={}\left(\tilde{P}/P_0\right)^{\frac{1}{2}}/r,
\end{equation} 
the mean rate of return  per period relative to $r.$
\begin{table}

\ra{1.3}
\centering
\begin{tabular}{@{}llll@{}}
\toprule
$(\alpha,\,\beta)$ & $\alpha\beta$ & min A & max A\\
\hline
(4, 0.3)  & 1.2 & 0.8495 & 1.6034\\
(1.33, 0.8)  & 1.064 & 1.0491 & 1.0779\\
(3.01, 0.34)          & 1.0234 &  0.8208&  1.2824 \\
(2, 0.5)       & 1       &    0.9160       &  1.0954\\
(5, 0.16) & 0.8 &0.5166 & 1.5053\\
(2.01, 0.3) & 0.603 & 0.4996 & 0.7772 \\
\bottomrule
\end{tabular}

\vspace{7pt}

\caption{Minimum and maximum  values of the rate function $A$  in 
\eqref{amplification} over the range $k_1,k_2\in[0.01, 100], $ $s_1,s_2\in [1,100],$ and  several pairs of values $\alpha$ and $\beta$ used in numerical simulations. \label{table:A}}
\end{table} Table \ref{table:A} shows numerically obtained min and max values of function $A$ for several values of $\alpha$ and $\beta.$ 
When $\min A>1$ the stock price grows every two periods at the rate strictly greater than the equilibrium rate $r^2.$ If $\min A\leq 1,$ the rate can be greater or less than the equilibrium rate, but the numerical simulations show that, after a short transient period, the growth rate over two periods  settles at a stationary value of 
\begin{equation}
\label{rate:stock}
r_s{}={}r(\alpha\beta)^{\frac{1}{2}},
\end{equation}
per period, with non-diminishing oscillations around it. The formula holds for wide range of values of the initial data and $(\alpha,\beta).$ It holds, for example for all pairs of values $(\alpha,\beta)$ presented in Table \ref{table:A}.  Formula \eqref{rate:stock} is the quantitative expression of how the adaptive behavior of agents is reflected in the stock price. In this model, the mean return is proportional to the geometric mean of rate of  change of the risk attitude.

\begin{figure}[t]
\centering
\includegraphics[width=7cm]{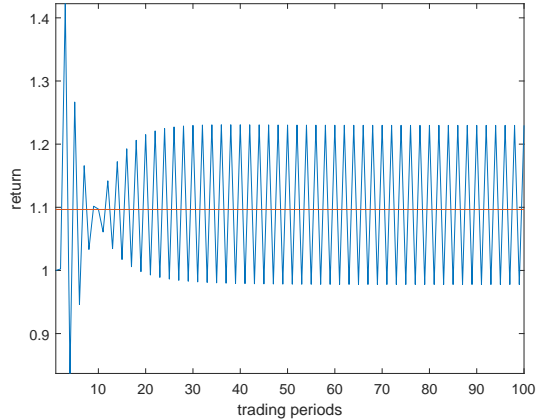}
\caption{The rate of gross return for 2-agent model: blue line is the ratio $P_n/P_{n-1},$ red line -- the return $r_s$ from \eqref{rate:stock}.  The return on the bond $0.01\%$ and stock returns $9.7\%$ per period. The plot is obtained for the values: $
r=1.001,\, \alpha=4,\,\beta = 0.3, \,k_1=k_1^s+0.01,\,k_2=k_2^s-0.01,\,s_i=s_i^s,\, b_i=b_i^s,\, P_0=1,
$
where $(k_i^s,s_i^s,b_i^s)$ is a steady state
$
k_1^s=0.5,\,k_2^s=0.8,\,b_1^s=b_2^s=10,\, s_i^s=k_i^sb_i^s.
$}
\label{fig:2-Return}
\end{figure}

\end{section}

\begin{section}{N-agent stochastic model}
\label{sec:3}

We consider a population of $N$ agents under the same conditions as in 2-agent model.  This time, $m$ agents are selected at random every trading period to set the new stock price $P$ through the demand-supply balance and to update their risk ratios. The asynchronous action of agents reflects the continuous flow of ask and bid orders in a trading system. It was used, for example, in the agent based model of Kim-Markowitz \cite{KM}.

Let $\{i_l\,:\,l=1..m\}$  be the set of  ``active'' agents, i.e. the ones setting the new price. If $\$x_{i_l}$ are the dollar amount that agent $i_l$ wants to invest, then
\[
\frac{\frac{P}{P_0}s_{i_l}+x_{i_l}}{rb_{i_l}-x_{i_l}}{}={}k_{i_l},
\] 
where, as before, $(k_i,s_i,b_i)$ is the state vector of the risk ratio, stock and bond investments for agent $i.$ The demand-supply balance is
\[
\sum_{l=1}^m x_{i_l}{}={}0,
\]
which can be solved for $P:$
\[
\frac{P}{P_0}{}={}r\left( \sum_{l=1}^m\frac{k_{i_l}b_{i_l}}{1+k_{i_l}}\right)\left(\sum_{l=1}^m\frac{s_{i_l}}{1+k_{i_l}}\right)^{-1}.
\]
The update mechanism is given by \eqref{update}. The interaction is repeated the following trading periods with randomly selected sets of active agents. As in the 2-agent model, there is a steady-state solution, when all agents have balanced portfolios, $s_i/b_i=k_i,$ $i=1,\ldots,N,$ and stock price grows at the rate $r:$ $P_{n}=rP_{n-1}.$ If the initial data are out of the steady-state, the system will exhibit non-trivial dynamics, diverging from the steady-state. We will proceed heuristically to derive the formula for the mean return of the stock. Following formula \eqref{rate:stock} we assume that the rate of return $r_s$ equals the product of rate of return on bond, $r,$ and geometric mean of the change of stock-to-bond ratio of agents per trading period. The probability that an agent is selected as ``active'' is the fraction $\frac{m}{N}.$ Assuming that his or her changes in the riskiness are equally likely, the geometric mean equals $(\alpha\beta)^{\frac{m}{2N}},$ and we get
\begin{equation}
\label{rate:stock_N}
r_s{}={}r(\alpha\beta)^{\frac{m}{2N}}.
\end{equation}
The formula is a good approximation of the mean return computed numerically, see Table \ref{tab:rate}.
\begin{table}
\centering
\begin{tabular}{@{}lrrrrr@{}}
\toprule
$m$                               &  $5$      & $10$   & $20$    & $40$    & $80$\\
\hline
Numerical Mean Returns  & 1.001   & 1.001  & 1.0016 & 1.00281 & 1.00536 \\
Analytical Mean Returns  & 1.0006 & 1.0006 & 1.0015 & 1.00283 & 1.00542 \\
\bottomrule
\end{tabular}

\vspace{7pt}

\caption{Mean for the distribution of returns and empirical formula
\eqref{rate:stock_N}. \label{tab:rate}}

\end{table}

\end{section}

\begin{figure}[t]
\centering
\includegraphics[width=14cm]{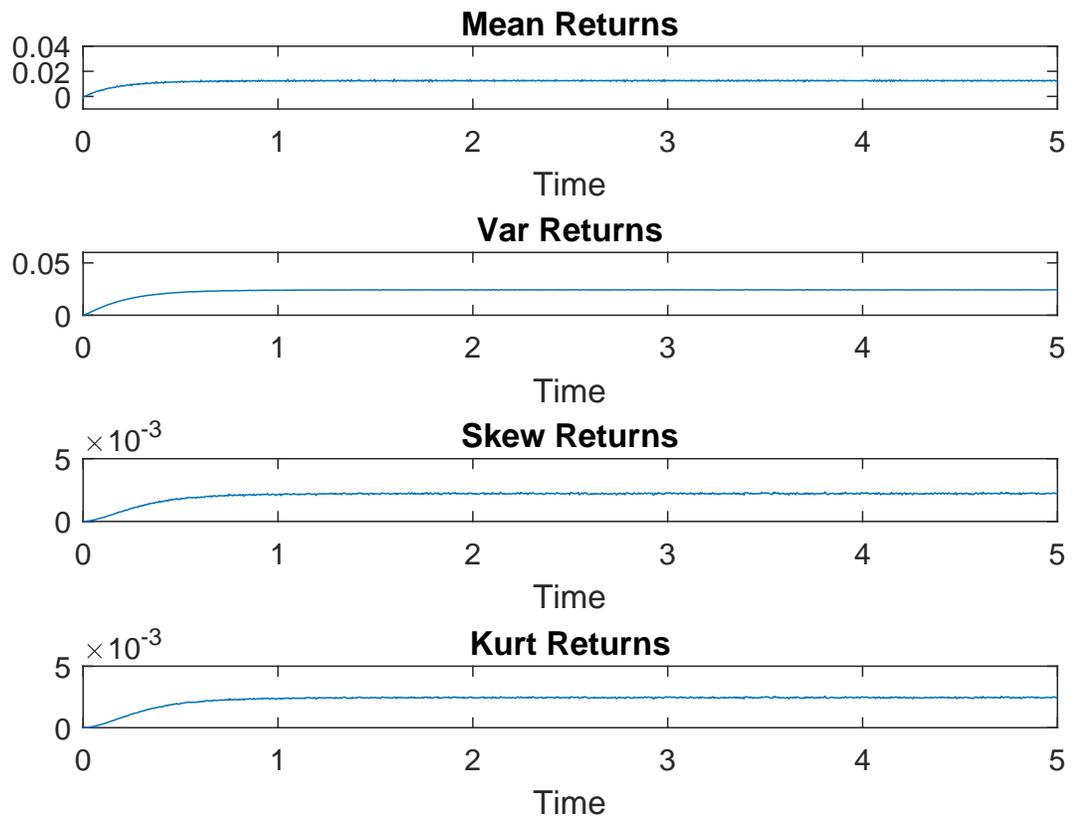}
\caption{Moments of Returns in Time; after time $t\approx 1$year the statistics of returns becomes stationary.}
\label{fig:mom}
\end{figure}

\begin{figure}[htbp]
   \centering
       \includegraphics[width=7cm]{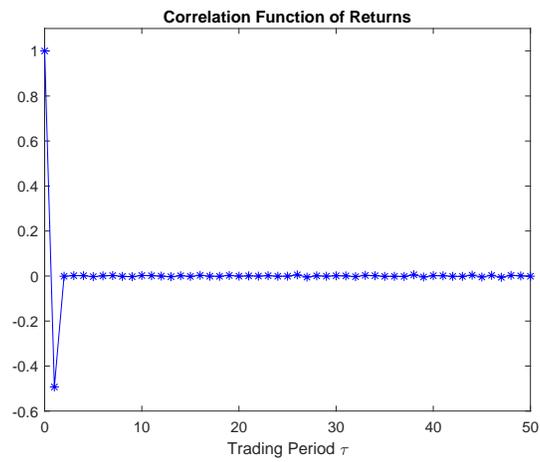}
   \caption{Correlation between  $\log R_n$ and $\log R_{t+\tau},$ where
   $n=2$years and $\tau$ is the trading period. Correlation for other 
   values of $n>2$ exhibits the same behavior due to the stationarity of returns.}
   \label{fig:cf}
\end{figure}

\begin{figure}[t]
\centering
\includegraphics[width=6cm]{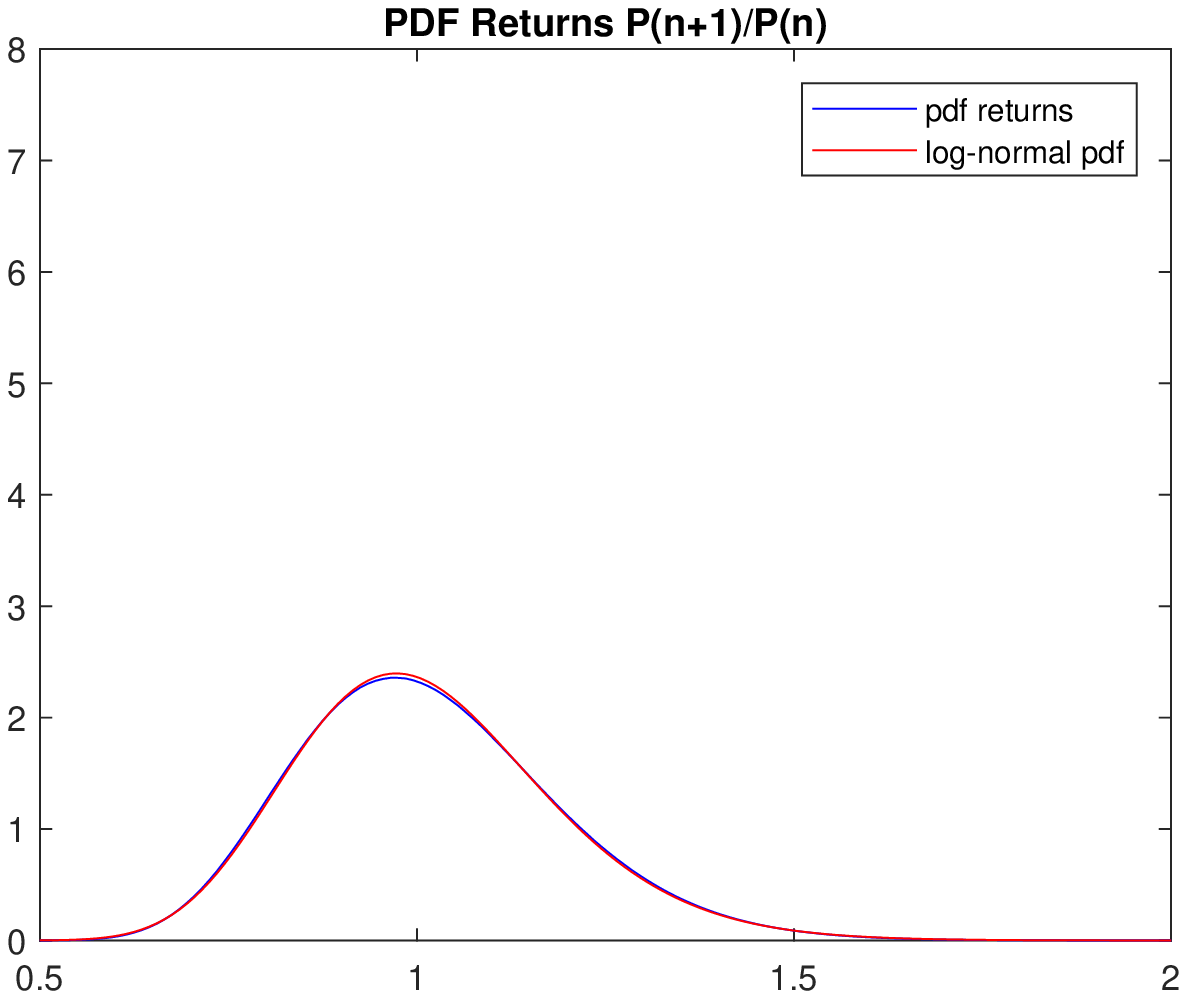}
\includegraphics[width=6cm]{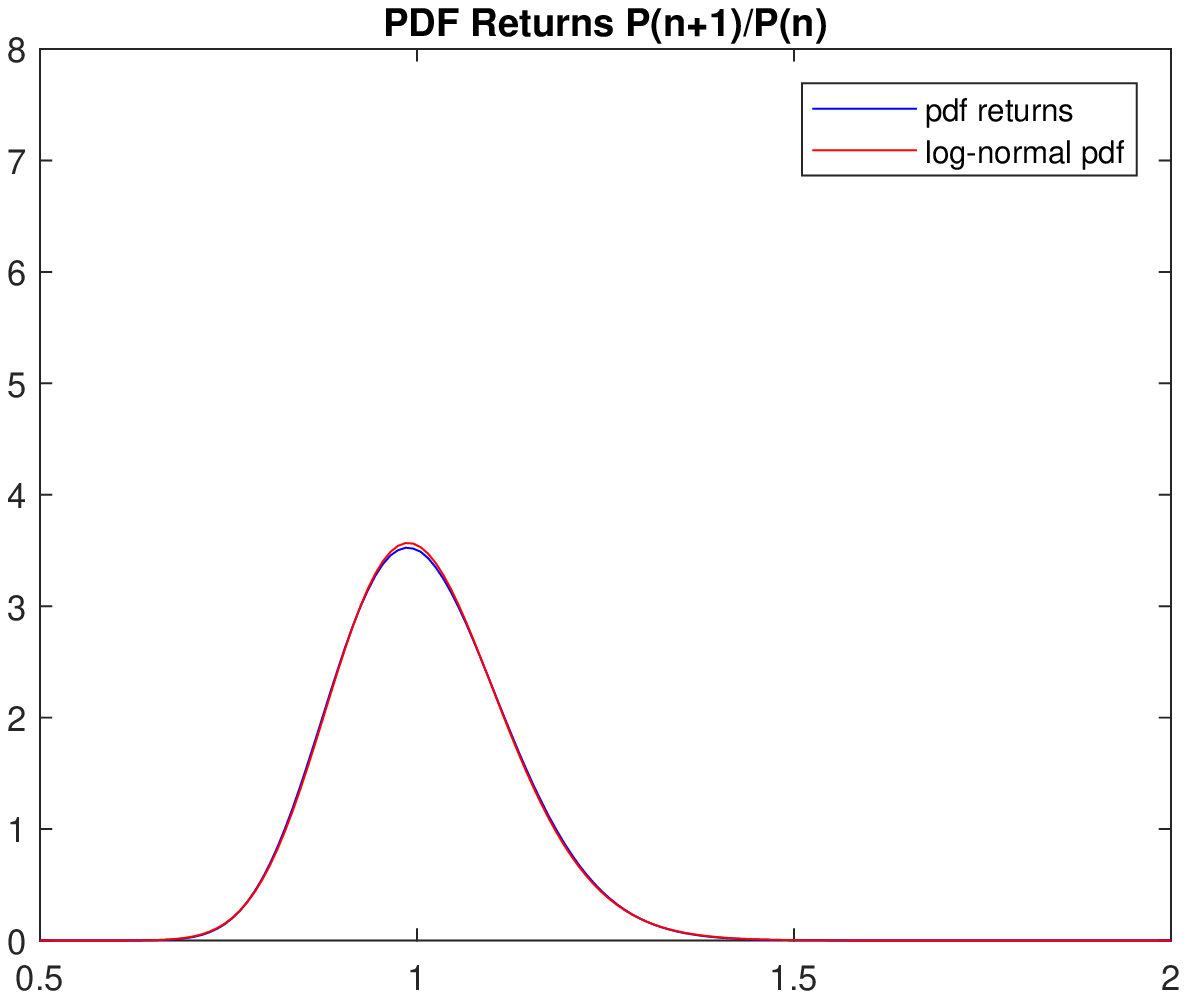}\\
\includegraphics[width=6cm]{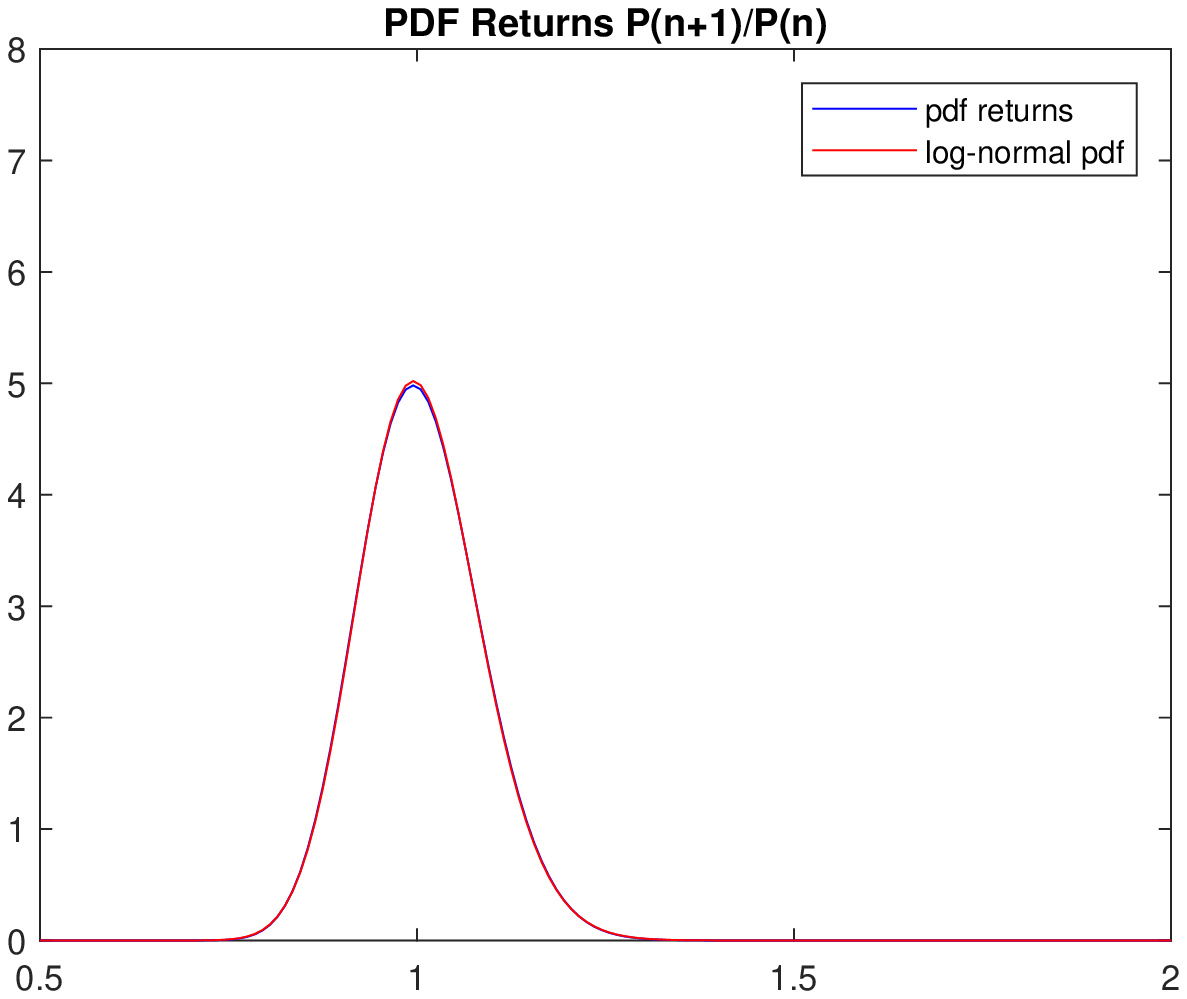}
\includegraphics[width=6cm]{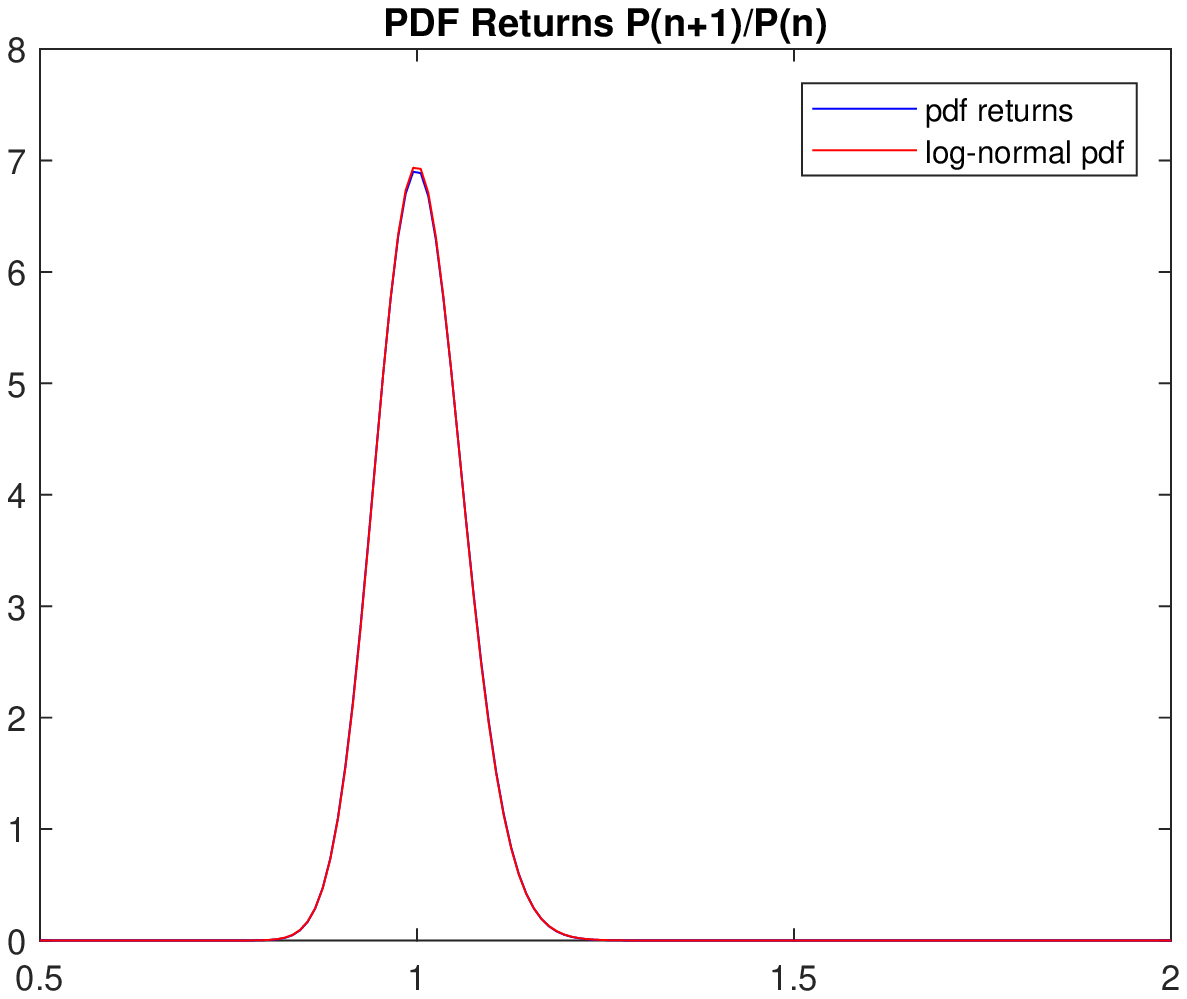}
\caption{Histogram of Returns computed from $MC=200,000$ trajectories and a fixed number of interacting agents $m=5$, $10$, $20$, $40$ and comparison with the log-normal density with the same mean and variance.}
\label{fig:pdf2}
\end{figure}

\begin{figure}[t]
\centering
\includegraphics[width=6cm]{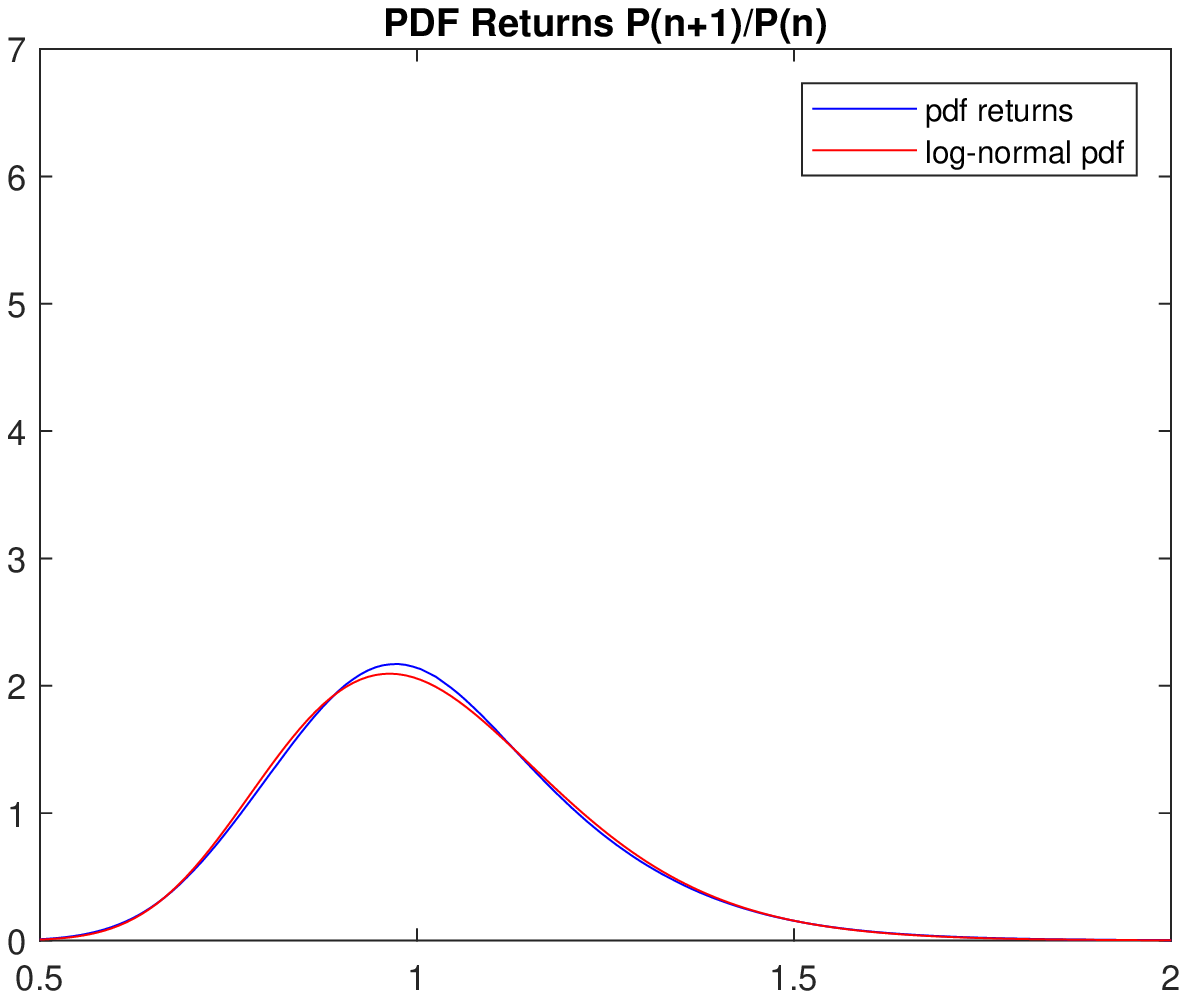}
\includegraphics[width=6cm]{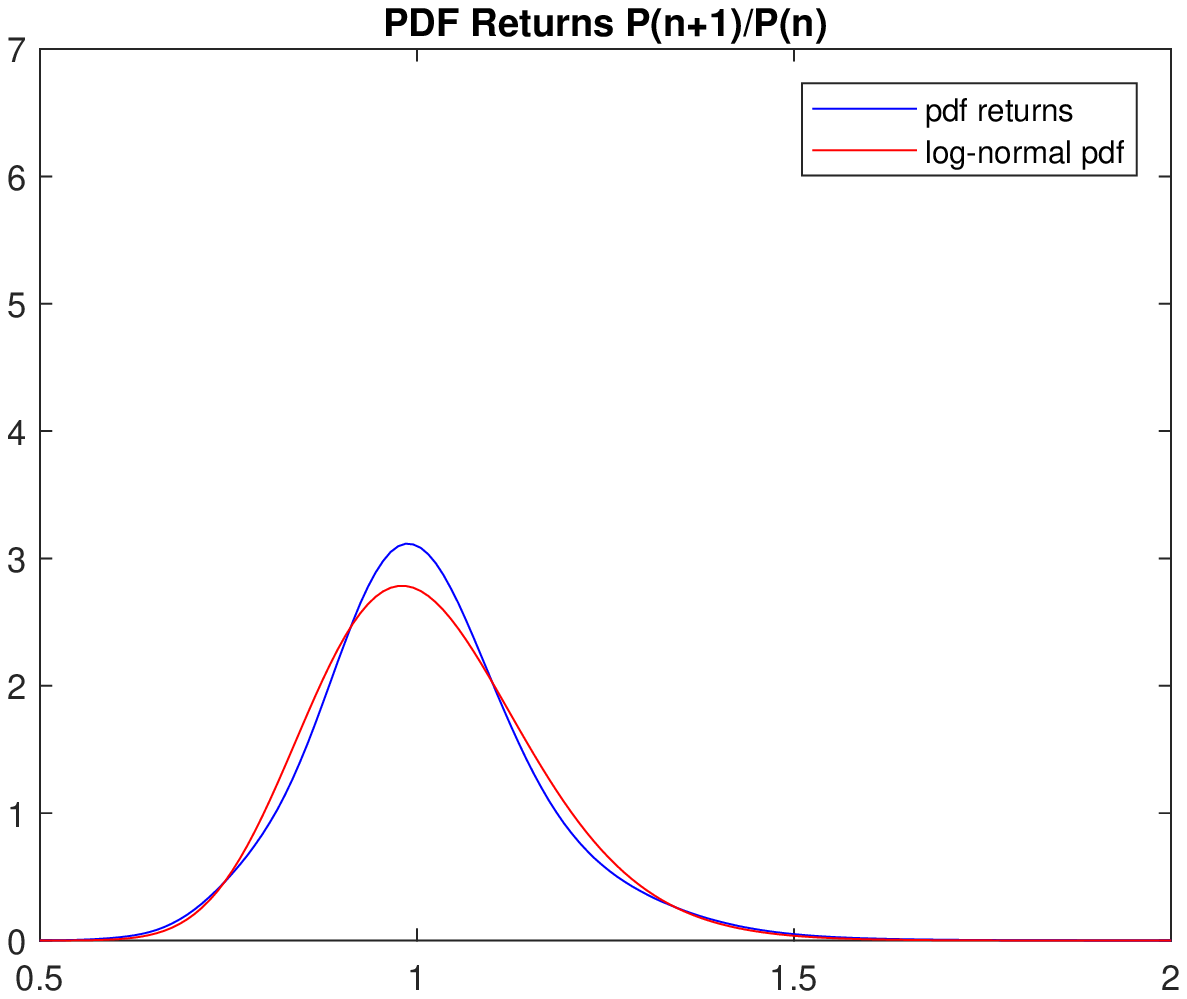}\\
\includegraphics[width=6cm]{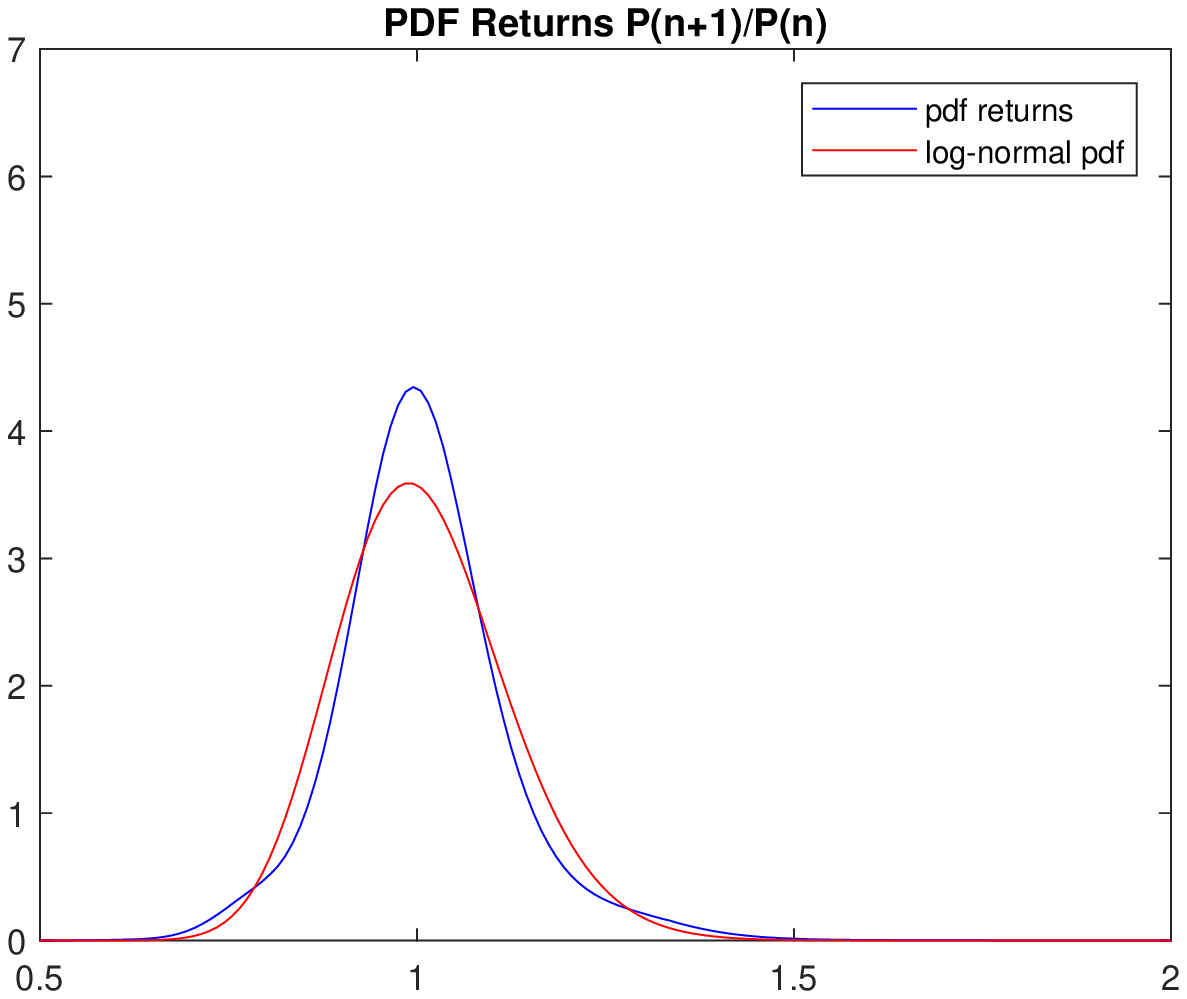}
\includegraphics[width=6cm]{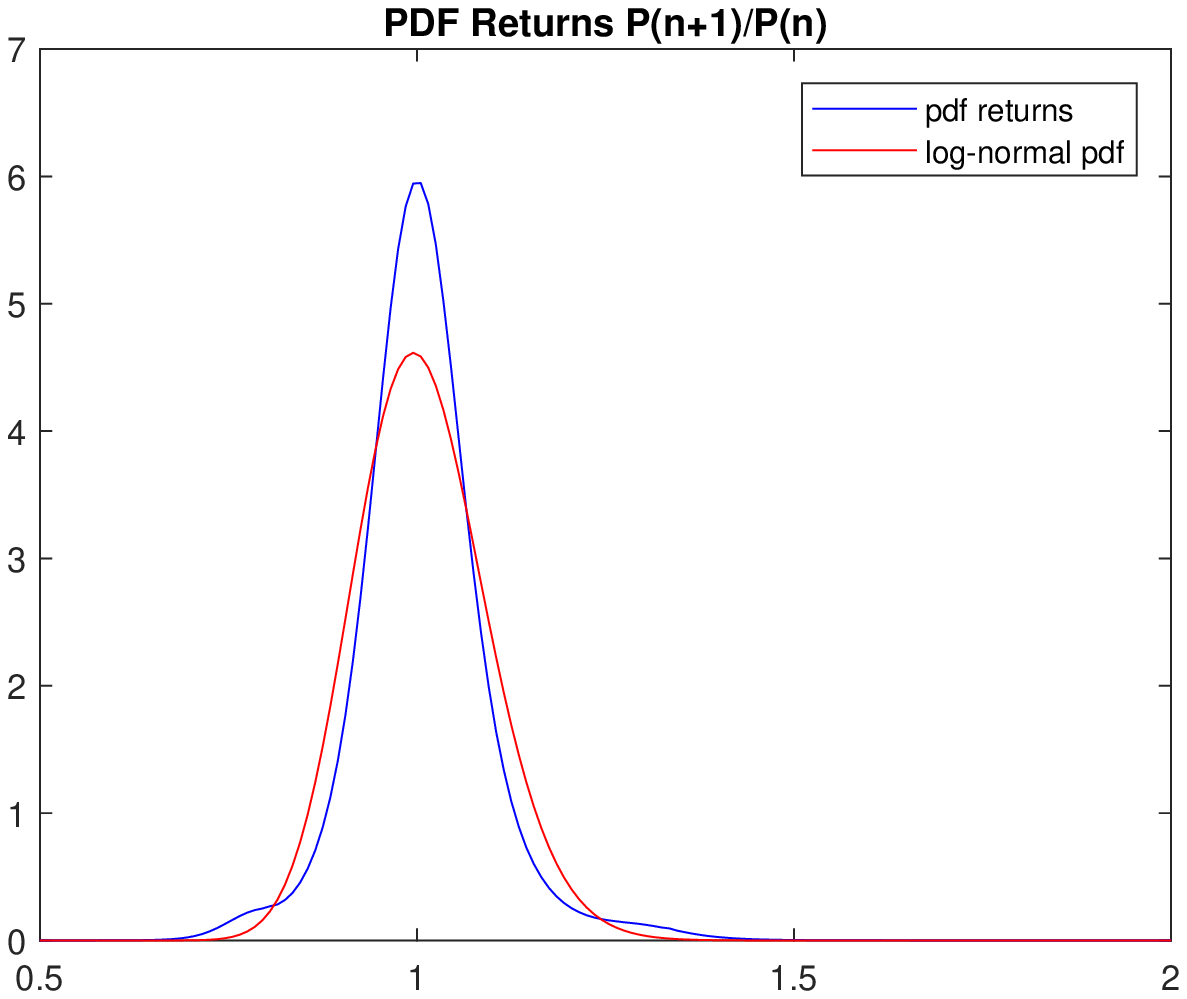}
\caption{Histogram of Returns computed from $MC=200,000$ trajectories and a random number of interacting agents with $m={\rm unif}(2,m_{max})$, $m_{max}=9$, $19$, $39$, $79$ so that the mean number of interacting agents is $\bar{m}=4.89$, $9.95$, $19.97$, $39.99$ which agrees roughly with the simulations
with a fixed number of agents depicted in Figure \ref{fig:pdf2}. Comparison with the log-normal density with the same mean and variance is also depicted for each case.}
\label{fig:pdf2a}
\end{figure}

\begin{section}{Numerical simulations}
\label{sec:4}
The purpose of this section is to confirm numerically the validity of formula \eqref{rate:stock_N} and to describe other statistical properties of the stochastic process with a large number of agents.  In particular we are interested in the distribution of returns, temporal correlation of returns, the distribution of stock and bond investments in the population of agents. 

In the simulation we take $N=500$ agents with a small number of active agents $m=5, 10, 20, 40,$ or $m$ is randomly selected from the uniform distribution over $2,\ldots,m_{max}$ so that the mean of this distribution approximately agrees with the 
fixed number of agents. The time scale is one year, during which a large number of trading periods take place. In particular, we set the number of trading periods per year to $M=200$ to mimic the daily trading sessions.
The average number of times an agent is selected as an active agent during a year  is $mM/N.$  
The simulations are run for $10$ years and start with  initial data in the vicinity of a steady-state solution. The time period of 10 years is of the order of the scale at which macroeconomic parameters change, and it is reasonable to assume that an agent adheres to a fixed strategy during that period, but might change it in longer runs.

We simulate $MC=200,000$ trajectories and depict the moments of the returns in
Figure \ref{fig:mom}. After time $t\approx1$ year the statistical behavior of returns becomes stationary. Thus, we skip $t=2$ years to compute the stationary properties of the returns in the model. The autocorrelation function with lags from 0 to 50 trading period is shown in Figure \ref{fig:cf}.
The next trading period return is negatively correlated with the current return, but the returns on the successive trading sessions appear to be uncorrelated. The negative correlation reflects the fact that large returns are recognized by most of the  agents as under- or over-performing markets and lead to the backward adjustment.

The histograms of returns are displayed in Figures \ref{fig:pdf2} and \ref{fig:pdf2a}. 
We compare the numerical histogram with the log-normal density with the same
mean and variance. When the number of active traders per period is fixed, $m=5,10, 20,40$, the return fits accurately to the corresponding log-normal distribution. With $m$ randomly chosen for each interacting period 
from interval ${\rm unif}(2,m_{max})$ with $m_{max}=9, 19, 39, 79$, the distribution has  sizable deviations from the log-normal distribution for larger values of $m_{max}$, see Figure \ref{fig:pdf2a}. 
The empirical histogram is narrower than the log-normal distribution.
The tails of the empirical distribution are noticeably heavier, especially for larger values of $m_{max}$. 

In addition, we also performed numerical simulations where the
number of interacting agents is randomly chosen from the binomial distribution with
$N=500$ and the ``probability of participating'' $p=0.05$, $0.1$, $0.2$, $0.4$.
When the number of interacting agents, $m$, is chosen from the binomial distribution
the distribution of returns agrees very well with the log-normal distribution (not depicted here). 
When comparing the binomial and the uniform distribution with the same means, 
the variance of the binomial is much smaller than that of the uniform distribution.
Therefore, we conjecture that the variance of the distribution of 
interacting agents is crucial in producing significant deviations from the 
log-normal distribution for the returns.

The simulated mean return is in good agreement with the formula \eqref{rate:stock_N}. Table \ref{tab:rate} compares the predicted and simulated means for a range of active traders. 
\end{section}

\begin{section}{Conclusions}

We showed that a group of traders, where each trader acts strictly on the basis of performance of his or her ``market'' portfolio, can create a positive price dynamics, with the mean return exceeding the return on a safe account. The value of the mean return and the degree of price oscillations around the mean are determined by the parameters of the model. The important characteristic of this trading process is stability. The numerical simulations show that the process moves to a stationary regime in which the mean return and variance do not grow in time. 

The analysis suggests the adaptive feedback trading as a plausible explanation
for stock indexes staying above the safe return rate when the economic data, such as dividends and earnings, stay flat.  The price dynamics generated by this process  can be described as a price bubble. After all, there is only one exogenous parameter - the return on a bond account. With no future dividends on a risky asset its fundamental value must be zero. The exponential growth of the price is  solely due to the adaptive trading mechanism.  Assuming that the efficient market hypothesis applies to the market environment in which traders operate, the trading process must be accompanied by growing systematic risk of the crash of the bubble, in a manner of rational bubbles described by Blanchard-Watson \cite{BW}. For a low or moderate growth rate of the asset price,  the systematic risk is not likely to be factored into the traders' decision making, and the bubble can persist for a long time, before ending in a crash.


In this paper we analyzed the properties of the stochastic process generated by adaptive trading. By the numerical simulations we fount that the process quickly moves to a regime in which  stock returns are stationary, while other state variables are not. When the number of interacting agents is fixed and does not change from one interacting period to another,
the distribution of returns has a log-normal distribution and negligible temporal correlations. 
However, when the number of interacting agents per period is random, the distribution of returns exhibits  
tails which are heavier than those of the log-normal. This is especially true when the 
distribution for the number of interacting agents has a relatively large variance. 
Therefore, the model exhibits rich statistical features for
the distribution of returns.

\end{section}

\clearpage

%
%
%

\end{document}